\documentclass[fleqn,usenatbib]{mnras}

\usepackage{newtxtext,newtxmath}

\usepackage[T1]{fontenc}
\usepackage{ae,aecompl}

\usepackage{graphicx}	
\usepackage{amsmath}	
\usepackage{gensymb}

\usepackage{scalerel}
\usepackage{tikz}
\usepackage{mathtools}

\usetikzlibrary{svg.path}

\definecolor{orcidlogocol}{HTML}{A6CE39}
\tikzset{
  orcidlogo/.pic={
    \fill[orcidlogocol] svg{M256,128c0,70.7-57.3,128-128,128C57.3,256,0,198.7,0,128C0,57.3,57.3,0,128,0C198.7,0,256,57.3,256,128z};
    \fill[white] svg{M86.3,186.2H70.9V79.1h15.4v48.4V186.2z}
                 svg{M108.9,79.1h41.6c39.6,0,57,28.3,57,53.6c0,27.5-21.5,53.6-56.8,53.6h-41.8V79.1z M124.3,172.4h24.5c34.9,0,42.9-26.5,42.9-39.7c0-21.5-13.7-39.7-43.7-39.7h-23.7V172.4z}
                 svg{M88.7,56.8c0,5.5-4.5,10.1-10.1,10.1c-5.6,0-10.1-4.6-10.1-10.1c0-5.6,4.5-10.1,10.1-10.1C84.2,46.7,88.7,51.3,88.7,56.8z};
  }
}

\newcommand\orcidicon[1]{\href{https://orcid.org/#1}{\mbox{\scalerel*{
\begin{tikzpicture}[yscale=-1,transform shape]
\pic{orcidlogo};
\end{tikzpicture}
}{|}}}}

\title[On the rise times in FU Orionis events]{On the rise times in FU Orionis events}

\author[Borchert et al.]{
Elisabeth M. A. Borchert\orcidicon{0000-0002-6994-8874},$^{1}$\thanks{E-mail: elisabeth.borchert@monash.edu}
Daniel J. Price\orcidicon{0000-0002-4716-4235},$^{1}$
Christophe Pinte\orcidicon{0000-0001-5907-5179},$^{1,2}$
Nicol\'as Cuello\orcidicon{0000-0003-3713-8073}$^{2}$
\\
$^{1}$School of Physics and Astronomy, Monash University, Clayton Vic 3800, Australia\\
$^{2}$Univ. Grenoble Alpes, CNRS, IPAG / UMR 5274, F-38000 Grenoble\\
}

\date{Accepted XXX. Received YYY; in original form ZZZ}

\pubyear{2021}

\begin{document}
\label{firstpage}
\pagerange{\pageref{firstpage}--\pageref{lastpage}}
\maketitle

\begin{abstract}
    We examine whether stellar flyby simulations can initiate FU Orionis outbursts using 3D hydrodynamics simulations coupled to live Monte Carlo radiative transfer. We find that a flyby where the secondary penetrates the circumprimary disc triggers a 1--2 year rise in the mass accretion rate to $10^{-4}~{\rm M_{\odot}~ yr^{-1}}$ that remains high ($\gtrsim 10^{-5}~{\rm M_{\odot}~yr^{-1}}$) for more than a hundred years, similar to the outburst observed in FU Ori. Importantly, we find that the less massive star becomes the dominant accretor, as observed.
\end{abstract}



\section{Introduction}
The year was 1936. Over the course of the year, a previously unremarkable star in Orion brightened by 6 magnitudes. FU Ori has remained bright ever since. Many explanations have been proposed, but none completely explain the phenomenon.

FU Ori is now the leading head of a class of low mass, pre-main sequence stars which experience sudden increase in brightness over a short period of time \citep{Herbig66a,Herbig77a}. Since the first detection, a number of objects have been classed as FU Orionis objects (FUors) based upon similar observed outbursts, or their spectral characteristics (see review by \citealt{Hartmann96a}). The outbursts, while yet to be explained, are thought to be an important aspect of the star formation process, leading to sudden bursts in the accretion rate \citep{Hartmann96a}.

Reasonable explanations for the outbursts range from disc thermal instability \citep{Lin85a,Clarke90a,Bell95a} to binary interactions \citep{Bonnell92a, Reipurth04a}, or perhaps a combination of the two. Binary interaction seems probable in FU Ori, in particular after \cite{Wang04a} first resolved FU Ori N and S. The binary was characterised further by \citet{Reipurth04a} and \citet{Beck12a}. More recent observations in scattered light \citep{Liu16a, Takami18a} and most recently with the Atacama Large Millimetre/Submillimetre Array (ALMA) \citep{Perez20a} show strong perturbations in the disc, suggesting past binary interactions, though other interpretations are possible (cf. \citealt{Takami18a}). Similar disturbed morphologies have been observed with ALMA around other FUors \citep[][]{Hales15a}  but have been interpreted as conical outflows \citep[][]{Zurlo17a, Ruiz-Rodriguez17a}.

In this letter we consider the hypothesis of FU Ori-like outbursts being induced by a stellar flyby \citep{Pfalzner08a, Cuello19a, Cuello20a, Vorobyov21a}. Flyby scenarios have previously been discounted due to a paucity of detected close companions (1/3; \citealt{Green16a})
as well as their inability to maintain outbursts for 100 years \citep{Clarke90a}. However, companions are easily obscured by disturbed material.

Importantly, observations have shown that in FU Ori the `primary' is not actually the primary. FU Ori N, the more luminous star, is the lower mass star (0.3--0.6 ${\rm M_{\odot}}$; \citealt{Zhu07a, Perez20a}). FU Ori N is the star in outburst, with $\dot{M}\approx 3.8\times 10^{-5}~{\rm M_{\odot}~yr^{-1}}$ \citep{Perez20a}. FU Ori S, the less luminous star, is the higher mass star (1.2 ${\rm M_{\odot}}$) and accretes at a normal rate, $\dot{M}\approx 2-3\times 10^{-8}~{\rm M_{\odot}~yr^{-1}}$ \citep{Beck12a}. We therefore should not be trying to trigger an outburst in the primary disc. Rather, \emph{the outburst needs to occur around the perturber}. \cite{Vorobyov21a} reproduced such outbursts with retro- and prograde disc-penetrating flybys, though not yet on timescales comparable to FUOrs.

Our main aim is to test whether a stellar flyby that penetrates the disc can produce a fast-rising but long-lasting outburst. We build on the flyby simulations performed by \cite{Cuello19a} but with on-the-fly live radiative transfer calculations. While our simulations are motivated by observations in FU Ori, we have not attempted a detailed reconstruction.

\begin{figure*}
    \centering
    \includegraphics[width=\textwidth]{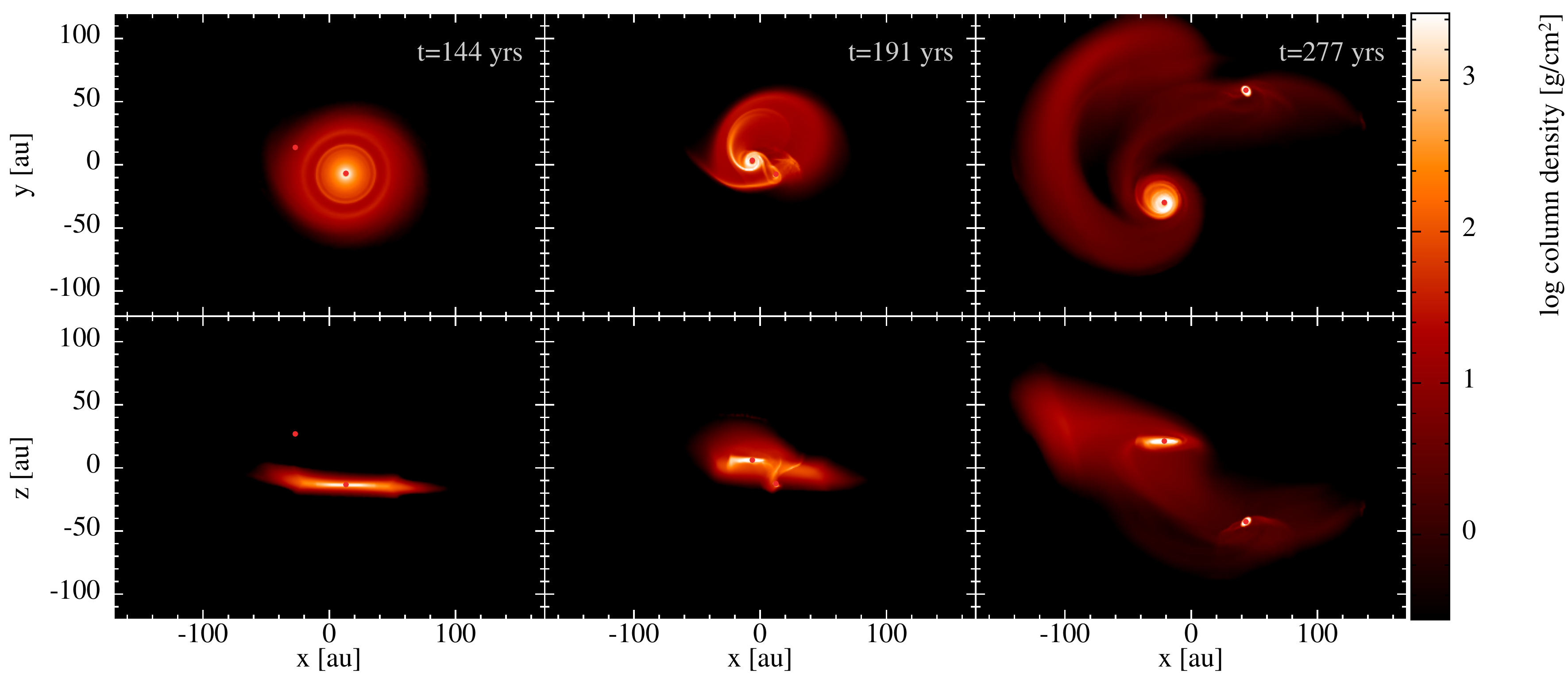}
    \caption{Column density of a stellar flyby at $r_{\rm P}=20$ au. The top row shows a top view and the bottom row a side view. Each column shows the same snap shots from the simulation, giving a visual representation $\sim35.5$ years before and $\sim11.5$ and $\sim97.5$ years after periastron (left to right). The red dots in the plots indicate the location of the stars. The secondary can be seen to grab material from the primary disc which becomes a circumsecondary disc.}
    \label{fig:density}
\end{figure*}

\section{Methods}\label{sec:methods}

We performed 3D radiation hydrodynamic simulations with the smoothed particle hydrodynamics (SPH) code \textsc{phantom} \citep{Price18b} coupled to the Monte Carlo radiative transfer code \textsc{mcfost} \citep{Pinte06a, Pinte09a}. Our simulations build on previous flyby simulations performed by \cite{Cuello19a} with two key differences. First, rather than prescribing a fixed radial temperature profile as in \cite{Cuello19a}, we updated the temperature live during the calculation by calling \textsc{mcfost} on-the-fly at set intervals. This allowed us to compute disc temperature profiles self-consistently, even if there were discs formed around both stars. Second, we included a contribution from the mass accretion rate in the stellar luminosity used to irradiate the disc(s) during the temperature calculations in \textsc{mcfost}. This allowed for the possibility of thermal feedback from the enhanced accretion rate induced by the flyby. We assumed that the gas and dust temperatures were equal when updating the temperature from \textsc{mcfost} in the SPH calculations. This assumption is valid except in the low density upper atmosphere \citep[e.g.][]{Woitke2009a}, and does not impact our estimated accretion rates. Our method implicitly assumes radiative equilibrium between the stellar radiation and the gas at all times.

We used the same setup as in \cite{Cuello19a}, using a prograde flyby at an angle of $\beta=45\degree$ (as this showed the highest accretion outburst), and flip the whole system for presentation in order to match observed disc rotation in FU Ori. The mass of the primary was set to $1.2~{\rm M_{\odot}}$ and the perturber mass to $0.6~{\rm M_{\odot}}$, based on observational estimates in FU Ori \citep{Zhu07a, Perez20a}. Stars were treated as sink particles with an accretion radius of 0.5 au. Their initial separation was set to ten times the periastron distance on a parabolic orbit. The primary was setup with a surrounding disc of $0.02~{\rm M_{\odot}}$, an inner radius of 1 au and an outer radius at 50 au, and an initial surface density profile \mbox{$\Sigma(r)\propto r^{-1}$}. For the purposes of placing the particles we assumed an initial $H/R=0.05(R/1~{\rm au})^{1/4}$, though this is free to change based on the temperature obtained from the radiative transfer code. As usual we use the SPH shock viscosity to mimic a disc viscosity, \cite{Lodato10a}, with $\alpha_{\rm av}=0.22$ giving $\alpha_{\rm disc}\approx 0.005$, appropriate to a weakly ionised disc. Our simulations were performed with $10^6$ SPH particles, initially in the primary disc.

We performed a set of simulations, changing the periastron distance while keeping the above parameters fixed. We used distances ranging from 5 au to 50 au in 5 au increments plus two additional simulations at 60 and 100 au. The time interval to call \textsc{mcfost} was set to $1/100$ of the time it takes the stellar flyby to reach the same separation past periastron as in the initial conditions at ten times periastron (computed using Barker's equation; \citealt{Cuello19a}), i.e. 100 calls over the course of the simulation --- 50 on the approach to periastron, and 50 after.

For each sink particle we computed the time averaged mass accretion rate $\dot{M}$ since the previous call to \mbox{\textsc{mcfost}}. Using this, we calculated the accretion luminosity \citep{Shakura73a} and temperature as
\begin{equation}
    L_{\rm acc}=\frac{1}{2} \frac{GM_{\star} \dot{M}}{R_{\star}}\ \ \ \mathrm{and}\ \ \  T_{\rm acc}=\left(\frac{L_{\rm acc}}{4\pi\sigma R_{\star}^2}\right)^{1/4},
\end{equation}
with the Stefan Boltzmann constant $\sigma$. The stellar radii $R_{\star}$ and effective temperatures are calculated in \textsc{mcfost} from the sink particle masses $M_{\star}$ using the 3~Myr isochrone from \cite{Siess2000}, which leads to $R_{\star}=1.4~{\rm R_{\odot}}$ and \mbox{$T_{\rm eff}=3882~{\rm K}$}. We then selected the closest stellar spectra in the "NextGen" database \citep{Hauschildt1999}. The accretion luminosity is emitted uniformly from the stellar surface (on top of the stellar luminosity) with a blackbody spectrum with $T=T_{\rm acc}$.

\begin{figure*}
    \centering
    \includegraphics[width=\textwidth]{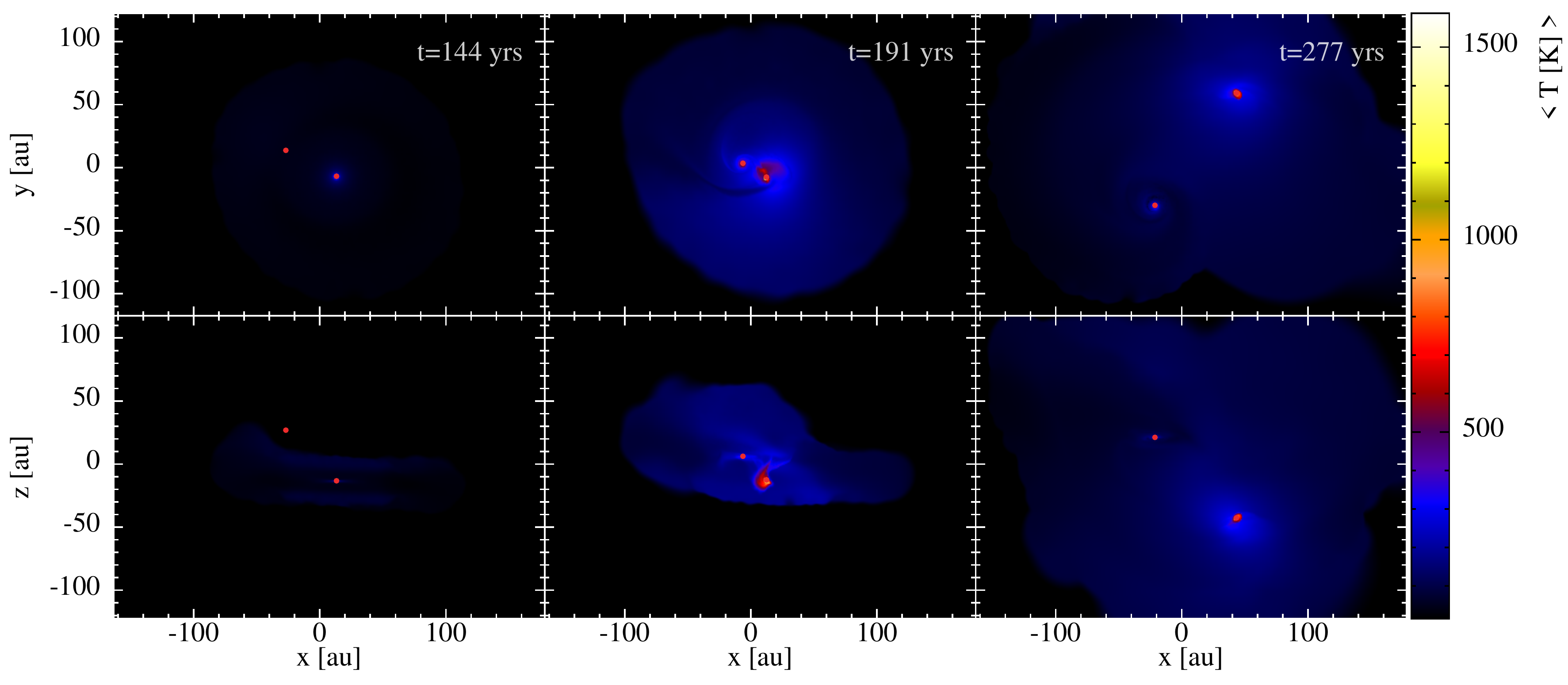}
    \caption{Normalised temperature plot for the same views and snapshots as in Figure~\ref{fig:density}. The highest normalised temperatures reach $\sim 1580$ K. After the perturber passes through the disc, the right column shows that the circumsecondary disc remains hotter than the disc surrounding the primary.}
    \label{fig:temperature}
\end{figure*}

\section{Results}\label{sec:results}

Figure~\ref{fig:density} shows column density in our simulation with \mbox{$r_{\rm P} = 20$ au}, 35.5 years before and 11.5 and 97.5 years after periastron (left to right, respectively) and from two perspectives (top and bottom rows, respectively). The rings in the primary disc are an artefact from the initial response to the inner disc boundary condition.

Figure~\ref{fig:temperature} shows density normalised temperature maps (line-of-sight integral of $\rho T$ divided by line-of-sight integral of $\rho$) for the same snapshots as in Figure~\ref{fig:density}. During the time of closest approach, temperatures around the perturber can be seen to reach up to \mbox{$\sim 1580$ K}. The maximum temperatures occur around the perturber during the process of penetrating the disc (middle panel).

\begin{figure}
    \centering
    \includegraphics[width=\columnwidth]{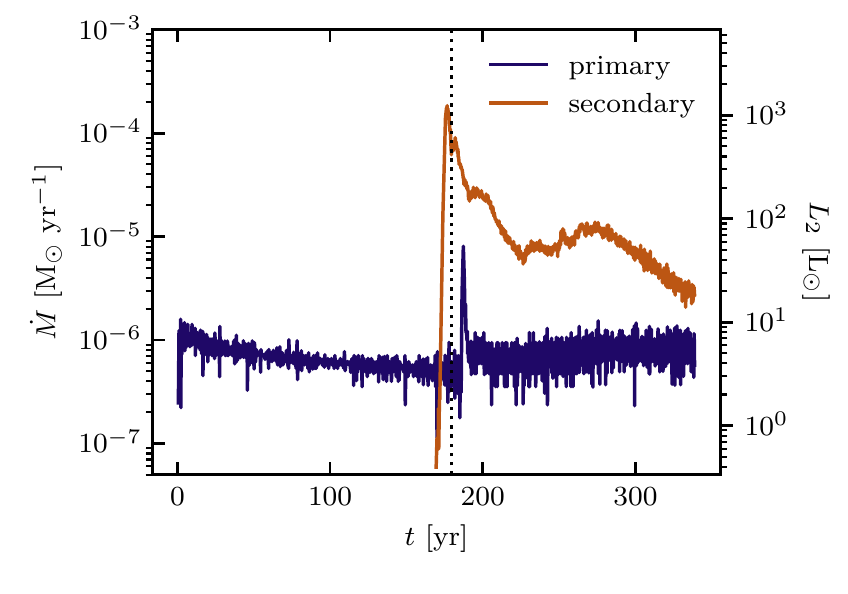}
    \caption{Time evolution of the accretion rate of the primary (dark blue) and the secondary (orange) from the simulation with $r_{\rm P}=20$ au. The right axis shows the accretion luminosity of the secondary. Dotted line indicates periastron at $\approx$179.5 years. We see an outburst in the secondary (happening from 171-175 yr) while the primary only experiences a small burst. Furthermore, the accretion rate, while not maintained at the maximum, continues at 1--2 orders of magnitude higher than what was observed pre-flyby in the primary.}
    \label{fig:mdot}
\end{figure}

While travelling through the primary disc, the secondary experiences a sudden spike in mass accretion, shown by the orange line in Figure~\ref{fig:mdot}. This rise begins when the secondary passes through the primary disc and captures material (bottom panel of Figure~\ref{fig:density}). Before interacting with the disc, our secondary does not accrete mass as there is no disc surrounding it. We observe a fast rise of the mass accretion rate onto the secondary in all simulations where the secondary penetrates the disc. The amplitude of the outburst decreases with increasing periastron distance. At \mbox{$r_{\rm P}=60$ au}, when the perturber no longer penetrates the primary disc, there remains a small accretion burst on the secondary, though of a considerably smaller magnitude and duration compared to penetrating encounters (less than half an order of magnitude increase in $\dot{M}_2$ over the pre-burst $\dot{M}_1$ at \mbox{$r_{\rm P}=60$ au} compared to $\sim$ 2 orders of magnitude for closer perturbers). The only simulations where we observed no outburst on the secondary used a periastron distance of 100 au. At such a distance the secondary disturbs the primary disc but does not capture a significant amount of material, giving $\dot{M}_2\lesssim 10^{-8}$M$_\odot/$yr.

\begin{figure}
    \centering
    \includegraphics[width=\columnwidth]{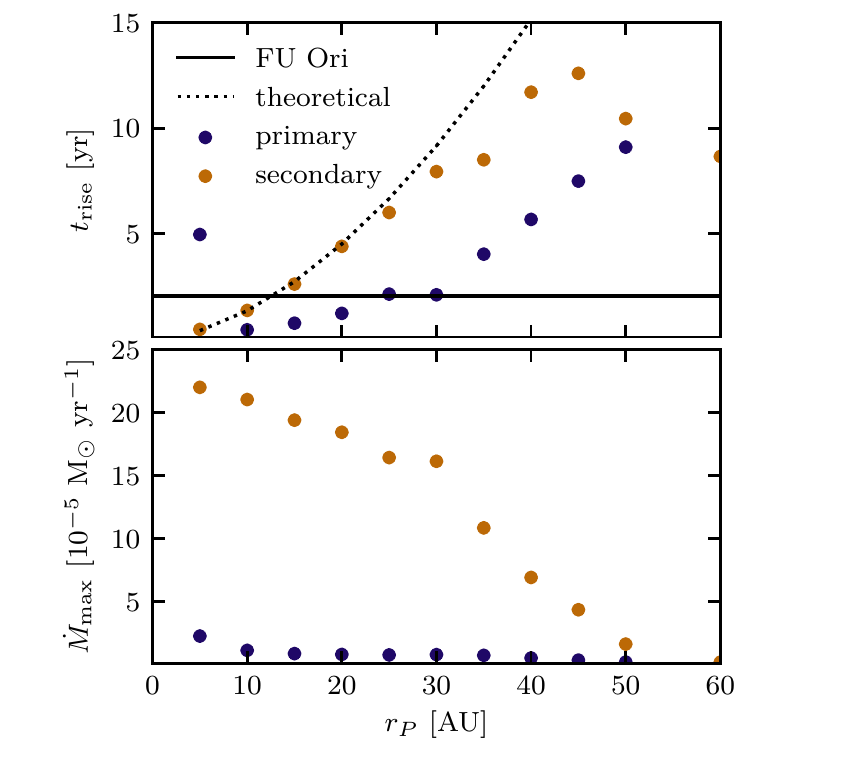}
    \caption{Rise time (top) and maximum mass accretion rate (bottom) for different periastron distances. Dark blue is for the primary star while orange is for the secondary. Horizontal line shows the rise time we calculated for FU Ori from its lightcurve \protect\citep[see Figure 3 in][]{Hartmann96a} in the top. The dotted line shows a theoretical estimate for the rise time of the stellar flyby following Equation~\eqref{eq:risetime}.}
    \label{fig:risetime}
\end{figure}

The dark blue line in Figure~\ref{fig:mdot} shows that, as the secondary experiences an accretion burst while passing through the disc, the primary also experiences a small burst in mass accretion, visible as a small spike at about 190 yr (11.5 years post-pericentre). This burst is slightly delayed compared to the burst in the secondary, and with a smaller amplitude. The primary outburst is caused by shocks in the disc excited by the perturber, which travel inward to create the burst in accretion on the primary. Up to periastron distances of \mbox{$r_{\rm P}=40$ au}, the amplitude of the outburst in the primary sits at about an order of magnitude above the pre-flyby mass accretion rate. The amplitude of the mass accretion rate for the primary decreases for more distant flybys and disappears completely when the secondary does not penetrate the disc.

Figure~\ref{fig:mdot} shows that $\dot{M}$ of the secondary rises by more than two orders of magnitude over a short time interval (rise time of a few years). This is an order of magnitude shorter than the orbital timescale of the primary disc at the perturber location (\mbox{$\approx80$ years}). 

In the top panel of Figure~\ref{fig:risetime} we show the rise times of different simulations as a function of periastron distance for our set of simulations. The bottom panel shows the maximum mass accretion rate of the outburst as a function of periastron distance. We defined the rise time as the time it takes for the accretion rate to reach the maximum accretion rate from the average accretion rate of the primary before the outburst. That is, the time it takes for the perturber's accretion rate to go from the primary's pre-burst accretion rate ($\sim5\times10^{-7}~{\rm M_{\odot}}$) to the peak accretion ($\sim2\times10^{-4}~{\rm M_{\odot}}$). The dark blue shows the rise time in the primary and the orange the rise time in the secondary. We included a horizontal line showing the rise time we measured the same way from the observed FU Ori lightcurve (Figure 3 in \citealt{Hartmann96a}). In general, we see a trend of longer rise times for increasing periastron distances. For \mbox{$r_{\rm P} \ge 35$ au} we observe a drop in the rise times of the secondary from the previously observed trend. At this point the amplitude of the outburst reduces from $\ge 2$ orders of magnitude to $< 2$ orders of magnitude.

We can understand the rapid timescale as follows: While the primary receives its accretion outburst from inward evolution of disc disturbances, the secondary goes into outburst when penetrating the disc of the primary, capturing material and accreting rapidly due to direct cancellation of angular momentum. We model the secondary outburst with a theoretical relation by assuming the perturber at speed $v=\sqrt{2GM/r_{\rm P}}$ passes through a certain distance $L(r_{\rm P})$ of the disc, shown as the dotted line in Figure~\ref{fig:risetime}. This leads to the theoretical rise time of:
\begin{equation}
    t_{\rm rise}=\frac{L(r_{\rm P})}{\sqrt{2GM/r_{\rm P}}},
    \label{eq:risetime}
\end{equation}
with the total mass of both stars $M$. The length $L(r_{\rm P})$ used for the theoretical line in Figure~\ref{fig:risetime} corresponds to $L(r_{\rm P})=5.7H$, with $H=r_{\rm P} \times (H/R)_{R=r_{\rm P}}$. Given the primary disc aspect ratio and the flaring index, this leads to the following dependence between the rise time and the periastron distance:
\begin{equation}
    t_{\rm rise}\propto r_{\rm P}^{5/4}.
\end{equation}
This relation holds when the perturber passes through the bulk of the disc. We can see in Figure~\ref{fig:risetime} that the results start to deviate significantly when the periastron distance approaches the outer edge of the disc. Figure~\ref{fig:risetime} shows that deviations from the theoretical trend occur for periastron distances above 30 au due to tidal truncation.

The accretion burst seen in Figure~\ref{fig:mdot} is sustained for more than 100 years by ongoing infall onto the circumsecondary disc from the surroundings.

\section{Discussion}\label{sec:discussion}
The main objection to binary interactions as the origin for FU Orionis outbursts relates to the fast rise time. According to \cite{Hartmann96a}: \emph{`To explain the fastest rise times of a year, the eruption must involve disc regions smaller than one au [because disc evolution will occur on timescales much longer than an orbital period \citep{Pringle81a}]'}. This makes the false assumption that the primary disc goes into outburst. This is not what was observed in FU Ori, where the more luminous star is actually the low mass star, i.e. the perturber \citep{Wang04a, Reipurth04a, Beck12a, Perez20a}.

In our simulations, it is indeed the secondary that goes into a sustained outburst. The luminosity of our secondary exceeds $100~{\rm L_{\odot}}$ for 40 years before continuing at about $40~{\rm L_{\odot}}$. The primary does \emph{not} go into a sustained outburst, although a spike in $\dot{M}_{1}$ occurs shortly after periastron. This also agrees with observations: FU Ori S, the more massive star, accretes at a normal rate with an unexceptional luminosity of $\sim2-3~{\rm L_{\odot}}$ \citep{Beck12a}.

The rise times of our outbursts are on a time scale that is comparable to the fastest timescales observed in FUors, e.g. in FU Ori and V1057 Cyg \citep[see Table 1 in][]{Hartmann96a}. This is visible in Figure~\ref{fig:risetime}, where we reproduce the rise time of FU Ori for a prograde stellar flyby at 10--15 au periastron distance. With our secondary going into outburst, we do not require the interaction to happen at a distance of less than 1 au. The main requirement for the secondary to erupt is that the secondary penetrates the primary disc at periastron. The trend seen in Figure~\ref{fig:mdot} is in line with the findings in \cite{Vorobyov21a}, except their rise time is of order \mbox{$\sim1$ kyr}, longer than those associated with FU Orionis outbursts. Closer flybys result in a faster and higher rise in the mass accretion rate. Wider separation passages may explain longer rise times in other FUors (e.g. V1515 Cyg). Our findings and the results from \cite{Vorobyov21a} show that flybys can trigger either fast or slow accretion bursts depending on whether or not the perturber penetrates the disc, respectively.

Our models of the secondary going into outburst may also explain the distinctive optical and near infrared characteristics of FUors \citep{Hartmann96a, Beck12a, Connelley18a}. First, they all have reflection nebulae \citep{Herbig77a, Goodrich87a}. Second, in near infrared FUors look like K-M giants while optical observations suggest effective temperatures of \mbox{6500--7200 K} and low surface gravities. We can explain this because our secondary is a low mass star (K-M) with a very high surface temperature, up to $10^4$ K depending on our assumed radius.

ALMA observations of FU Ori revealed direct evidence for binary interactions \citep{Zurlo17a, Cieza18a, Perez20a, Kospal21a}. In particular, CO line observations have revealed distinct blue and red shifted velocities corresponding to expanding `shells or rings'. These flows, previously interpreted as conical outflows \citep{Zurlo17a,Ruiz-Rodriguez17a,Ruiz-Rodriguez17b}, are naturally explained by line of sight motions of our interacting discs post-periastron. Our models show both blue and redshifted line of sight motions of order $5-20~{\rm km~s^{-1}}$, see e.g. synthetic line observations in \cite{Cuello20a}.

Instead of binary interactions, disc instabilities have previously been the favoured scenario for FU Orionis outbursts \citep{Bell95a}. In our simulations we triggered an outburst on the observed timescale that is maintained for more than 100 years without the need for thermal instability. Figure~\ref{fig:temperature} shows the temperature evolution in our flyby scenario. Through the use of on-the-fly radiative transfer calculations we were able to model the changes in disc temperature. Around the time of periastron, temperatures in the circumsecondary disc increase significantly. The disc temperature reaches up to \mbox{$\sim1580$~K} close to the perturber. These temperatures, while high, are still considerably below the temperatures required to initiate thermal instability. We therefore demonstrate that FU Orionis outbursts are possible without the need for thermal instability.

While the simulations do not reach the temperatures required for thermal instability, we do observe temperatures rising above 1500 K (see middle column in Figure~\ref{fig:temperature}). Dust should sublimate at this point as the temperatures are above 1000 K. Preliminary tests where we tried removing dust at temperatures above 1000 K did not change our results and thus we did not account for this in our final models. Interestingly, such rapid heating of the disc with dust melting could possibly explain the existence of chondrules in our solar system, where dynamical evidence for a past flyby exists \citep{Pfalzner18a}. Flash heating occurs to some extent around both stars, so evidence for similar heating in the solar system does not necessarily imply that our Sun was the perturber.

In our simulations we have not attempted a detailed reconstruction of the close encounter in FU Orionis. This is the likely reason we are not yet able to maintain the outburst exactly as seen in FU Ori. We have cases though where the secondary maintains a higher mass accretion rate than that of the primary (at least an order of magnitude for more than 100 years). One such example is shown in Figure~\ref{fig:mdot}. Sustained outbursts occur when the secondary captures material to form a circumsecondary disc with subsequent ongoing infall. One such case is the simulation presented in Figures~\ref{fig:density} and \ref{fig:mdot}, with the perturber disc visible in the right column in Figure~\ref{fig:density}.

We did not include magnetic fields in our calculations. Doing so would enable us to launch $\gtrsim200~{\rm km/s}$ jets from the inner disc, as seen in several FUors (e.g. V1057 Cyg; \citealt{Levreault88a}).

Our simulations provide an explanation for single FU Orionis outburst events. As we assumed a penetrating unbound stellar flyby, our secondary will only pass through periastron once, leading to one outburst. There are FUors which have experienced multiple outbursts (e.g. Z CMa) which have previously been predicted by bound binary interactions \citep{Bonnell92a}.

\section{Conclusions}\label{sec:conclusion}

We performed a series of numerical experiments of thermal feedback in 3D simulations of stellar flybys. We have been able to reproduce three key features of the observed outburst in FU Ori.

\begin{enumerate}
    \item We were able to reproduce a fast accretion rise as seen in FU Ori, with perturbers that penetrate the disc (no thermal instability).
    \item The mass accretion rate of the secondary continues at a higher level (1--2 orders of magnitude) to what it had been pre-flyby for the primary for more than 100 years. This occurs because of ongoing infall from the environment onto the circumsecondary disc.
    \item Our simulations further show the observed phenomenon of the lower mass star going into outburst, as is the case in FU Ori.
    
\end{enumerate}

It would be interesting to explore whether different stellar flybys at different angles, with different disc properties and accretion radii or a disc around the perturber are able to sustain the outburst at a higher amplitude for longer. In preliminary attempts at disc-disc flybys we find more sustained outbursts depending on the relative initial disc orientations. Another important follow-up should be to create synthetic observations and lightcurves in order to reconstruct what was observed in FU Ori in 1937.

\section*{Acknowledgements}
We thank our referee for a constructive and timely report and Iain Hammond, Evgeni Grishin and Giuseppe Lodato for useful discussions. DJP thanks Agnes K\'osp\'al and Peter \'Abrah\'am for useful discussions following the ``Great Barriers in Planet Formation'' conference in 2019. EMAB acknowledges an Australian government Research Training Program (RTP) and a Monash International Tuition Scholarship. We acknowledge Australian Research Council grants FT170100040 and DP180104235. NC acknowledges European Union H2020 funding via Marie Skłodowska-Curie grant 210021. We used Ozstar and Gadi via the National Computing Infrastructure / Swinburne University. We used \textsc{splash} \citep{Price07a}, \textsc{Matplotlib} \citep{matplotlib}, \textsc{NumPy} \citep{numpy} and \textsc{CMasher} \citep{cmasher}. 

\vspace{-0.5cm}

\section*{Data availability}
\textsc{phantom}\footnote{\url{https://github.com/danieljprice/phantom}} and \textsc{mcfost}\footnote{\url{https://github.com/cpinte/mcfost}} are publicly available
and the simulation setup and data underlying this article will be shared on request. 

\vspace{-0.5cm}

\bibliographystyle{mnras}
\bibliography{bib} 

\begin{thebibliography}{}
\makeatletter
\relax
\def\mn@urlcharsother{\let\do\@makeother \do\$\do\&\do\#\do\^\do\_\do\%\do\~}
\def\mn@doi{\begingroup\mn@urlcharsother \@ifnextchar [ {\mn@doi@}
  {\mn@doi@[]}}
\def\mn@doi@[#1]#2{\def\@tempa{#1}\ifx\@tempa\@empty \href
  {http://dx.doi.org/#2} {doi:#2}\else \href {http://dx.doi.org/#2} {#1}\fi
  \endgroup}
\def\mn@eprint#1#2{\mn@eprint@#1:#2::\@nil}
\def\mn@eprint@arXiv#1{\href {http://arxiv.org/abs/#1} {{\tt arXiv:#1}}}
\def\mn@eprint@dblp#1{\href {http://dblp.uni-trier.de/rec/bibtex/#1.xml}
  {dblp:#1}}
\def\mn@eprint@#1:#2:#3:#4\@nil{\def\@tempa {#1}\def\@tempb {#2}\def\@tempc
  {#3}\ifx \@tempc \@empty \let \@tempc \@tempb \let \@tempb \@tempa \fi \ifx
  \@tempb \@empty \def\@tempb {arXiv}\fi \@ifundefined
  {mn@eprint@\@tempb}{\@tempb:\@tempc}{\expandafter \expandafter \csname
  mn@eprint@\@tempb\endcsname \expandafter{\@tempc}}}

\bibitem[\protect\citeauthoryear{{Beck} \& {Aspin}}{{Beck} \&
  {Aspin}}{2012}]{Beck12a}
{Beck} T.~L.,  {Aspin} C.,  2012, \mn@doi [\aj] {10.1088/0004-6256/143/3/55},
  \href {https://ui.adsabs.harvard.edu/abs/2012AJ....143...55B} {143, 55}

\bibitem[\protect\citeauthoryear{{Bell}, {Lin}, {Hartmann}  \& {Kenyon}}{{Bell}
  et~al.}{1995}]{Bell95a}
{Bell} K.~R.,  {Lin} D.~N.~C.,  {Hartmann} L.~W.,   {Kenyon} S.~J.,  1995,
  \mn@doi [\apj] {10.1086/175612}, \href
  {https://ui.adsabs.harvard.edu/abs/1995ApJ...444..376B} {444, 376}

\bibitem[\protect\citeauthoryear{{Bonnell} \& {Bastien}}{{Bonnell} \&
  {Bastien}}{1992}]{Bonnell92a}
{Bonnell} I.,  {Bastien} P.,  1992, \mn@doi [\apjl] {10.1086/186663}, \href
  {https://ui.adsabs.harvard.edu/abs/1992ApJ...401L..31B} {401, L31}

\bibitem[\protect\citeauthoryear{{Cieza} et~al.,}{{Cieza}
  et~al.}{2018}]{Cieza18a}
{Cieza} L.~A.,  et~al., 2018, \mn@doi [\mnras] {10.1093/mnras/stx3059}, \href
  {https://ui.adsabs.harvard.edu/abs/2018MNRAS.474.4347C} {474, 4347}

\bibitem[\protect\citeauthoryear{{Clarke}, {Lin}  \& {Pringle}}{{Clarke}
  et~al.}{1990}]{Clarke90a}
{Clarke} C.~J.,  {Lin} D.~N.~C.,   {Pringle} J.~E.,  1990, \mn@doi [\mnras]
  {10.1093/mnras/242.3.439}, \href
  {https://ui.adsabs.harvard.edu/abs/1990MNRAS.242..439C} {242, 439}

\bibitem[\protect\citeauthoryear{{Connelley} \& {Reipurth}}{{Connelley} \&
  {Reipurth}}{2018}]{Connelley18a}
{Connelley} M.~S.,  {Reipurth} B.,  2018, \mn@doi [\apj]
  {10.3847/1538-4357/aaba7b}, \href
  {https://ui.adsabs.harvard.edu/abs/2018ApJ...861..145C} {861, 145}

\bibitem[\protect\citeauthoryear{{Cuello} et~al.,}{{Cuello}
  et~al.}{2019}]{Cuello19a}
{Cuello} N.,  et~al., 2019, \mn@doi [\mnras] {10.1093/mnras/sty3325}, \href
  {https://ui.adsabs.harvard.edu/abs/2019MNRAS.483.4114C} {483, 4114}

\bibitem[\protect\citeauthoryear{{Cuello} et~al.,}{{Cuello}
  et~al.}{2020}]{Cuello20a}
{Cuello} N.,  et~al., 2020, \mn@doi [\mnras] {10.1093/mnras/stz2938}, \href
  {https://ui.adsabs.harvard.edu/abs/2020MNRAS.491..504C} {491, 504}

\bibitem[\protect\citeauthoryear{{Goodrich}}{{Goodrich}}{1987}]{Goodrich87a}
{Goodrich} R.~W.,  1987, \mn@doi [\pasp] {10.1086/131963}, \href
  {https://ui.adsabs.harvard.edu/abs/1987PASP...99..116G} {99, 116}

\bibitem[\protect\citeauthoryear{{Green} et~al.}{{Green}
  et~al.}{2016}]{Green16a}
{Green} J.~D.,  et~al., 2016, \mn@doi [\apj] {10.3847/0004-637X/830/1/29},
  \href {https://ui.adsabs.harvard.edu/abs/2016ApJ...830...29G} {830, 29}

\bibitem[\protect\citeauthoryear{{Hales} et~al.}{{Hales}
  et~al.}{2015}]{Hales15a}
{Hales} A.~S.,  et~al., 2015, \mn@doi [\apj] {10.1088/0004-637X/812/2/134},
  \href {https://ui.adsabs.harvard.edu/abs/2015ApJ...812..134H} {812, 134}

\bibitem[\protect\citeauthoryear{{Harris} et~al.,}{{Harris}
  et~al.}{2020}]{numpy}
{Harris} C.~R.,  et~al., 2020, \mn@doi [\nat] {10.1038/s41586-020-2649-2},
  \href {https://ui.adsabs.harvard.edu/abs/2020Natur.585..357H} {585, 357}

\bibitem[\protect\citeauthoryear{{Hartmann} \& {Kenyon}}{{Hartmann} \&
  {Kenyon}}{1996}]{Hartmann96a}
{Hartmann} L.,  {Kenyon} S.~J.,  1996, \mn@doi [\araa]
  {10.1146/annurev.astro.34.1.207}, \href
  {https://ui.adsabs.harvard.edu/abs/1996ARA&A..34..207H} {34, 207}

\bibitem[\protect\citeauthoryear{{Hauschildt}, {Allard}  \&
  {Baron}}{{Hauschildt} et~al.}{1999}]{Hauschildt1999}
{Hauschildt} P.~H.,  {Allard} F.,   {Baron} E.,  1999, \mn@doi [\apj]
  {10.1086/306745}, \href
  {https://ui.adsabs.harvard.edu/abs/1999ApJ...512..377H} {512, 377}

\bibitem[\protect\citeauthoryear{{Herbig}}{{Herbig}}{1966}]{Herbig66a}
{Herbig} G.~H.,  1966, \mn@doi [Vistas in Astronomy]
  {10.1016/0083-6656(66)90025-0}, \href
  {https://ui.adsabs.harvard.edu/abs/1966VA......8..109H} {8, 109}

\bibitem[\protect\citeauthoryear{{Herbig}}{{Herbig}}{1977}]{Herbig77a}
{Herbig} G.~H.,  1977, \mn@doi [\apj] {10.1086/155615}, \href
  {https://ui.adsabs.harvard.edu/abs/1977ApJ...217..693H} {217, 693}

\bibitem[\protect\citeauthoryear{{Hunter}}{{Hunter}}{2007}]{matplotlib}
{Hunter} J.~D.,  2007, \mn@doi [Computing in Science and Engineering]
  {10.1109/MCSE.2007.55}, \href
  {https://ui.adsabs.harvard.edu/abs/2007CSE.....9...90H} {9, 90}

\bibitem[\protect\citeauthoryear{{K{\'o}sp{\'a}l} et~al.,}{{K{\'o}sp{\'a}l}
  et~al.}{2021}]{Kospal21a}
{K{\'o}sp{\'a}l} {\'A}.,  et~al., 2021, \mn@doi [\apjs]
  {10.3847/1538-4365/ac0f09}, \href
  {https://ui.adsabs.harvard.edu/abs/2021ApJS..256...30K} {256, 30}

\bibitem[\protect\citeauthoryear{{Levreault}}{{Levreault}}{1988}]{Levreault88a}
{Levreault} R.~M.,  1988, \mn@doi [\apj] {10.1086/166520}, \href
  {https://ui.adsabs.harvard.edu/abs/1988ApJ...330..897L} {330, 897}

\bibitem[\protect\citeauthoryear{{Lin}, {Papaloizou}  \& {Faulkner}}{{Lin}
  et~al.}{1985}]{Lin85a}
{Lin} D.~N.~C.,  {Papaloizou} J.,   {Faulkner} J.,  1985, \mn@doi [\mnras]
  {10.1093/mnras/212.1.105}, \href
  {https://ui.adsabs.harvard.edu/abs/1985MNRAS.212..105L} {212, 105}

\bibitem[\protect\citeauthoryear{{Liu} et~al.,}{{Liu} et~al.}{2016}]{Liu16a}
{Liu} H.~B.,  et~al., 2016, \mn@doi [Science Advances]
  {10.1126/sciadv.1500875}, \href
  {https://ui.adsabs.harvard.edu/abs/2016SciA....2E0875L} {2, e1500875}

\bibitem[\protect\citeauthoryear{{Lodato} \& {Price}}{{Lodato} \&
  {Price}}{2010}]{Lodato10a}
{Lodato} G.,  {Price} D.~J.,  2010, \mn@doi [\mnras]
  {10.1111/j.1365-2966.2010.16526.x}, \href
  {https://ui.adsabs.harvard.edu/abs/2010MNRAS.405.1212L} {405, 1212}

\bibitem[\protect\citeauthoryear{{P{\'e}rez} et~al.,}{{P{\'e}rez}
  et~al.}{2020}]{Perez20a}
{P{\'e}rez} S.,  et~al., 2020, \mn@doi [\apj] {10.3847/1538-4357/ab5c1b}, \href
  {https://ui.adsabs.harvard.edu/abs/2020ApJ...889...59P} {889, 59}

\bibitem[\protect\citeauthoryear{{Pfalzner}}{{Pfalzner}}{2008}]{Pfalzner08a}
{Pfalzner} S.,  2008, \mn@doi [\aap] {10.1051/0004-6361:200810879}, \href
  {https://ui.adsabs.harvard.edu/abs/2008A&A...492..735P} {492, 735}

\bibitem[\protect\citeauthoryear{{Pfalzner}, {Bhandare}, {Vincke}  \&
  {Lacerda}}{{Pfalzner} et~al.}{2018}]{Pfalzner18a}
{Pfalzner} S.,  {Bhandare} A.,  {Vincke} K.,   {Lacerda} P.,  2018, \mn@doi
  [\apj] {10.3847/1538-4357/aad23c}, \href
  {https://ui.adsabs.harvard.edu/abs/2018ApJ...863...45P} {863, 45}

\bibitem[\protect\citeauthoryear{{Pinte}, {M{\'e}nard}, {Duch{\^e}ne}  \&
  {Bastien}}{{Pinte} et~al.}{2006}]{Pinte06a}
{Pinte} C.,  {M{\'e}nard} F.,  {Duch{\^e}ne} G.,   {Bastien} P.,  2006, \mn@doi
  [\aap] {10.1051/0004-6361:20053275}, \href
  {https://ui.adsabs.harvard.edu/abs/2006A&A...459..797P} {459, 797}

\bibitem[\protect\citeauthoryear{{Pinte} et~al.}{{Pinte}
  et~al.}{2009}]{Pinte09a}
{Pinte} C.,  et~al., 2009, \mn@doi [\aap] {10.1051/0004-6361/200811555}, \href
  {https://ui.adsabs.harvard.edu/abs/2009A&A...498..967P} {498, 967}

\bibitem[\protect\citeauthoryear{{Price}}{{Price}}{2007}]{Price07a}
{Price} D.~J.,  2007, \mn@doi [\pasa] {10.1071/AS07022}, \href
  {https://ui.adsabs.harvard.edu/abs/2007PASA...24..159P} {24, 159}

\bibitem[\protect\citeauthoryear{{Price} et~al.,}{{Price}
  et~al.}{2018}]{Price18b}
{Price} D.~J.,  et~al., 2018, \mn@doi [\pasa] {10.1017/pasa.2018.25}, \href
  {https://ui.adsabs.harvard.edu/abs/2018PASA...35...31P} {35, e031}

\bibitem[\protect\citeauthoryear{{Pringle}}{{Pringle}}{1981}]{Pringle81a}
{Pringle} J.~E.,  1981, \mn@doi [\araa] {10.1146/annurev.aa.19.090181.001033},
  \href {https://ui.adsabs.harvard.edu/abs/1981ARA&A..19..137P} {19, 137}

\bibitem[\protect\citeauthoryear{{Reipurth} \& {Aspin}}{{Reipurth} \&
  {Aspin}}{2004}]{Reipurth04a}
{Reipurth} B.,  {Aspin} C.,  2004, \mn@doi [\apjl] {10.1086/422250}, \href
  {https://ui.adsabs.harvard.edu/abs/2004ApJ...608L..65R} {608, L65}

\bibitem[\protect\citeauthoryear{{Ru{\'\i}z-Rodr{\'\i}guez}
  et~al.,}{{Ru{\'\i}z-Rodr{\'\i}guez} et~al.}{2017a}]{Ruiz-Rodriguez17a}
{Ru{\'\i}z-Rodr{\'\i}guez} D.,  et~al., 2017a, \mn@doi [\mnras]
  {10.1093/mnras/stw3378}, \href
  {https://ui.adsabs.harvard.edu/abs/2017MNRAS.466.3519R} {466, 3519}

\bibitem[\protect\citeauthoryear{{Ru{\'\i}z-Rodr{\'\i}guez}
  et~al.}{{Ru{\'\i}z-Rodr{\'\i}guez} et~al.}{2017b}]{Ruiz-Rodriguez17b}
{Ru{\'\i}z-Rodr{\'\i}guez} D.,  et~al., 2017b, \mn@doi [\mnras]
  {10.1093/mnras/stx703}, \href
  {https://ui.adsabs.harvard.edu/abs/2017MNRAS.468.3266R} {468, 3266}

\bibitem[\protect\citeauthoryear{{Shakura} \& {Sunyaev}}{{Shakura} \&
  {Sunyaev}}{1973}]{Shakura73a}
{Shakura} N.~I.,  {Sunyaev} R.~A.,  1973, \aap, \href
  {https://ui.adsabs.harvard.edu/abs/1973A&A....24..337S} {500, 33}

\bibitem[\protect\citeauthoryear{{Siess}, {Dufour}  \& {Forestini}}{{Siess}
  et~al.}{2000}]{Siess2000}
{Siess} L.,  {Dufour} E.,   {Forestini} M.,  2000, \aap, \href
  {https://ui.adsabs.harvard.edu/abs/2000A&A...358..593S} {358, 593}

\bibitem[\protect\citeauthoryear{{Takami} et~al.,}{{Takami}
  et~al.}{2018}]{Takami18a}
{Takami} M.,  et~al., 2018, \mn@doi [\apj] {10.3847/1538-4357/aad2e1}, \href
  {https://ui.adsabs.harvard.edu/abs/2018ApJ...864...20T} {864, 20}

\bibitem[\protect\citeauthoryear{{Vorobyov} et~al.}{{Vorobyov}
  et~al.}{2021}]{Vorobyov21a}
{Vorobyov} E.~I.,  et~al., 2021, \mn@doi [\aap] {10.1051/0004-6361/202039391},
  \href {https://ui.adsabs.harvard.edu/abs/2021A&A...647A..44V} {647, A44}

\bibitem[\protect\citeauthoryear{{Wang}, {Apai}, {Henning}  \&
  {Pascucci}}{{Wang} et~al.}{2004}]{Wang04a}
{Wang} H.,  {Apai} D.,  {Henning} T.,   {Pascucci} I.,  2004, \mn@doi [\apjl]
  {10.1086/381705}, \href
  {https://ui.adsabs.harvard.edu/abs/2004ApJ...601L..83W} {601, L83}

\bibitem[\protect\citeauthoryear{{Woitke}, {Kamp}  \& {Thi}}{{Woitke}
  et~al.}{2009}]{Woitke2009a}
{Woitke} P.,  {Kamp} I.,   {Thi} W.~F.,  2009, \aap, 501, 383

\bibitem[\protect\citeauthoryear{{Zhu} et~al.}{{Zhu} et~al.}{2007}]{Zhu07a}
{Zhu} Z.,  et~al., 2007, \mn@doi [\apj] {10.1086/521345}, \href
  {https://ui.adsabs.harvard.edu/abs/2007ApJ...669..483Z} {669, 483}

\bibitem[\protect\citeauthoryear{{Zurlo} et~al.,}{{Zurlo}
  et~al.}{2017}]{Zurlo17a}
{Zurlo} A.,  et~al., 2017, \mn@doi [\mnras] {10.1093/mnras/stw2845}, \href
  {https://ui.adsabs.harvard.edu/abs/2017MNRAS.465..834Z} {465, 834}

\bibitem[\protect\citeauthoryear{{van der Velden}}{{van der
  Velden}}{2020}]{cmasher}
{van der Velden} E.,  2020, \mn@doi [The Journal of Open Source Software]
  {10.21105/joss.02004}, \href
  {https://ui.adsabs.harvard.edu/abs/2020JOSS....5.2004V} {5, 2004}

\makeatother
\end{thebibliography}





\bsp	
\label{lastpage}
\end{document}